\documentclass[cits]{PoS}

\usepackage{amsmath,epsfig,latexsym,bm}

\title{Fluctuations and reweighting of the quark determinant on large 
lattices}

\ShortTitle{Quark determinant reweighting}

\author{\vspace{-0.3cm}
        \hfill{\rm CERN-PH-TH/2008-205}\\
        \vspace{0.3cm}}

\author{Martin L\"uscher, \speaker{Filippo Palombi}
        \\
        CERN, Physics Department, CH-1211 Geneva~23, Switzerland
        \\
        E-mail: 
        \email{Martin.Luescher@cern.ch},
        \email{Filippo.Palombi@cern.ch}
        }

\abstract{
We propose to stabilise HMC simulations of lattice QCD 
with very light Wilson quarks by splitting the quark determinant
into two factors and by treating the factor that includes
the contribution of the low modes of the Dirac operator
as a reweighting factor. In general,
determinant reweighting becomes inefficient on large lattices,
because the statistical fluctuations of quark
determinants increase exponentially with the lattice volume.
Random matrix theory and some numerical studies now suggest that the
low-mode contribution to the determinant behaves differently,
which allows factorisations to be devised that preserve the
efficiency of the simulation on large lattices.
}

\FullConference{The XXVI International Symposium on Lattice Field Theory \\
                 July 14 - 19, 2008\\
                 Williamsburg, Virginia, USA}

\begin{document}

% Definitions and abbreviations

% Roman letters in math formulae

\def\rmd{{\rm d}}
\def\rmD{{\rm D}}
\def\rme{{\rm e}}
\def\rmO{{\rm O}}
\def\rmU{{\rm U}}
\def\rmo{{\rm o}}

% Real and integer numbers

\def\rz{{\Bbb R}}
\def\gz{{\Bbb Z}}
\def\nz{{\Bbb N}}
\def\Im{{\rm Im}\,}
\def\Re{{\rm Re}\,}

% Special relations and symbols

\def\defeq{\mathrel{\mathop=^{\rm def}}}
\def\proof{\noindent{\sl Proof:}\kern0.6em}
\def\endproof{\hskip0.6em plus0.1em minus0.1em
\setbox0=\null\ht0=5.4pt\dp0=1pt\wd0=5.3pt
\vbox{\hrule height0.8pt
\hbox{\vrule width0.8pt\box0\vrule width0.8pt}
\hrule height0.8pt}}
\def\dual{\mathstrut^*\kern-0.1em}
\def\mod{\;\hbox{\rm mod}\;}
\def\ring{\mathaccent"7017}
\def\lvec#1{\setbox0=\hbox{$#1$}
    \setbox1=\hbox{$\scriptstyle\leftarrow$}
    #1\kern-\wd0\smash{
    \raise\ht0\hbox{$\raise1pt\hbox{$\scriptstyle\leftarrow$}$}}
    \kern-\wd1\kern\wd0}
\def\rvec#1{\setbox0=\hbox{$#1$}
    \setbox1=\hbox{$\scriptstyle\rightarrow$}
    #1\kern-\wd0\smash{
    \raise\ht0\hbox{$\raise1pt\hbox{$\scriptstyle\rightarrow$}$}}
    \kern-\wd1\kern\wd0}
\def\slash#1{\setbox0=\hbox{$#1$}\setbox1=\hbox{$\kern1pt/$}
    #1\kern-\wd0\kern1pt/\kern-\wd1\kern\wd0}

% Lattice derivatives

\def\nab#1{{\nabla_{#1}}}
\def\nabstar#1{{\nabla\kern0.5pt\smash{\raise 4.5pt\hbox{$\ast$}}
               \kern-5.5pt_{#1}}}
\def\drv#1{{\partial_{#1}}}
\def\drvstar#1{{\partial\kern0.5pt\smash{\raise 4.5pt\hbox{$\ast$}}
               \kern-6.0pt_{#1}}}
\def\ldrv#1{{\lvec{\,\partial}_{#1}}}
\def\ldrvstar#1{{\lvec{\,\partial}\kern-0.5pt\smash{\raise 4.5pt\hbox{$\ast$}}
               \kern-5.0pt_{#1}}}

% Units

\def\MeV{{\rm MeV}}
\def\GeV{{\rm GeV}}
\def\TeV{{\rm TeV}}
\def\fm{{\rm fm}}

% Constants

\def\euler{\gamma_{\rm E}}

% Fields

\def\psibar{\overline{\psi}}
\def\psitilde{\widetilde{\psi}}

% Dirac matrices

\def\dirac#1{\gamma_{#1}}
\def\diracstar#1#2{
    \setbox0=\hbox{$\gamma$}\setbox1=\hbox{$\gamma_{#1}$}
    \gamma_{#1}\kern-\wd1\kern\wd0
    \smash{\raise4.5pt\hbox{$\scriptstyle#2$}}}
\def\dirachat{\hat{\gamma}_5}

% Gauge group

\def\group{G}
\def\algebra{{\frak g}}
\def\SUtwo{{\rm SU(2)}}
\def\SUthree{{\rm SU(3)}}
\def\SUn{{\rm SU}(N)}
\def\tr{{\rm tr}}
\def\Tr{{\rm Tr}}
\def\Ad{{\rm Ad}\kern0.1em}

% Masses and decay constants

\def\mpi{M_{\pi}}
\def\Mpi{\mpi}
\def\msea{m_{\rm sea}}
\def\ZA{Z_{\rm A}}
\def\ZP{Z_{\rm P}}
\def\ksea{\kappa_{\rm sea}}
\def\kval{\kappa_{\rm val}}
\def\mur{\mu_{\rm R}}
\def\nub{\bar{\nu}_n}
\def\nubr{\bar{\nu}_{n,{\rm R}}}
\def\mr{m_{\rm R}}

% Parameters and abbreviations

\def\Dmod{\tilde{D}}
\def\nbrak#1{\langle{#1}\rangle}
\def\mbrak#1{\langle{#1}\rangle_{\rm m}}
\def\cbrak#1{\langle{#1}\rangle_{\rm con}}
\def\msbar{\overline{\rm MS\kern-1pt}\kern1pt}
\def\Neta{N}

\section{Introduction}

Simulations of lattice QCD with very light Wilson quarks
are potentially affected by algorithmic instabilities,
sampling inefficiencies and ergodicity violations.
The issue was studied in some detail in ref.~\cite{Stability} and 
it was concluded that simulations based on the Hybrid Monte Carlo 
(HMC) algorithm \cite{HMC} can be expected to be stable 
in a range of the lattice parameters and the 
light-quark masses which includes the
large-volume regime of QCD at lattice spacings $a\leq0.1$ fm.

The algorithm proposed here avoids the instabilities
from the beginning by separating the low modes of the Dirac
operator from the rest of the modes and by including 
only the latter in the HMC algorithm. The low modes are
then taken into account by reweighting the generated
representative ensemble of fields by the appropriate factor.
Similar mode separations were recently considered by 
Jansen et al.~\cite{tmQCDeps} and by
Hasenfratz et al.~\cite{MassRew}, 
partly with the same motivations and partly for other reasons
(see refs.~\cite{TDAI,TDAII,SFRew} for related
earlier work).

Low-mode reweighting tends to reduce
the statistical fluctuations of observables that are sensitive
to the low modes \cite{SFRew} but will only work out if
the reweighting factor itself does not fluctuate too much.
From this point of view,
reweighting by ratios of quark determinants
(such as the ones considered below) does not seem to be 
particularly promising,
because quark determinants scale exponentially
with the volume of the lattice. 
Our aim in this report is to show
that the situation is actually more favourable than
suspected, the main reason being that the
low eigenvalues of the Dirac operator (except perhaps for the few
lowest ones) fluctuate
by no more than a distance inversely proportional to 
the lattice volume about their mean values.

\section{Determinant factorisation}

We consider lattice QCD with a doublet of light Wilson quarks
and any number of heavier quarks. The (massive) light-quark
Wilson--Dirac operator is denoted by $D$ and the associated 
bare current-quark mass by $m$. On average the spectral 
gap of the hermitian Dirac operator $\dirac{5}D$ around the origin 
is then approximately equal to $\ZA m$, where $\ZA$ is the renormalization
constant of the isovector axial current \cite{Stability}.

Determinant reweighting starts from an exact factorisation
\begin{equation}
  \det(D^{\dagger}D)=W\det(\Dmod^{\dagger}\Dmod)
\end{equation}
of the light-quark determinant, where $W$ is the reweighting factor and
$\Dmod$ a modified Wilson--Dirac operator whose determinant is 
included in the HMC algorithm. 
In the following, two choices
$\Dmod_l$, $l=1,2$, of the modified operator will be considered
for which the associated reweighting factors are of the form
\begin{equation}
  W_l=\det\{w_l(D^{\dagger}D)\}
\end{equation}
(see Table~1). Note that the 
(complex) spectrum of $\dirac{5}\Dmod_l$ is
in both cases rigorously separated from the origin by 
a distance of order $\mu$, for all quark masses $m$,
very much as in twisted-mass QCD with twisted mass $\mu$ 
\cite{tmQCD}. The
modified quark determinant $\det(\Dmod_1{}\kern-2pt^{\dagger}\Dmod_1)$
in fact coincides with the quark determinant in twisted-mass QCD.

\begin{table}[t]
\small
\centering
%Blanke Zahl
%\newdimen\digitwidth
%\setbox0=\hbox{\rm 0}
%\digitwidth=\wd0
%\catcode`@=\active
%\def@{\kern\digitwidth}
\renewcommand{\arraystretch}{1.5}
\tabcolsep0.6cm
\begin{tabular}{cccc}
$l$ &
$\Dmod_l$ & 
$w_l(\nu^2)$ & 
$\left.w_l(\nu^2)\right|_{\nu^2\gg \mu^2}$\\[0.8ex]
\hline\hline
\noalign{\vskip1.3ex}
   $1$ &
   $\displaystyle D+i\mu\dirac{5}$ &
   $\displaystyle \frac{\nu^2}{\nu^2+\mu^2}$ &
   $\displaystyle 1-\frac{\mu^2}{\nu^2}+\rmO(\nu^{-4})$ \\[3ex]
   $2$ &
   $\displaystyle (D+i\mu\dirac{5})\frac{\dirac{5}D-i\mu}
   {\dirac{5}D-i\sqrt{2}\mu}$ &
   $\displaystyle \frac{\nu^2(\nu^2+2\mu^2)}{(\nu^2+\mu^2)^2}$ &
   $\displaystyle 1-\frac{\mu^4}{\nu^4}+\rmO(\nu^{-6})$ \\
\noalign{\vskip1.3ex}
\hline\hline
\noalign{\vskip0.2ex}
\end{tabular}
\caption{%
Modified Dirac operators and associated reweighting factors 
considered in this report. The mass parameter $\mu>0$
can in principle be set to any value, but 
good reweighting efficiencies are only achieved if
$\mu$ is not much larger than $\ZA m$.
}
\label{tab1}
\end{table}

The inclusion of the modified quark
determinant in the HMC algorithm is not expected
to give rise to instabilities since the modified Wilson--Dirac
operator is safe from having near-zero modes.
Moreover, we do not foresee any difficulties in applying 
acceleration techniques
such as the Schwarz preconditioning \cite{DDHMC}
and local deflation \cite{Deflation} to the modified algorithm.
In this report, however, the focus will be on the
reweighting efficiency and its dependence on the lattice size.

\section{Statistical fluctuations of $\bm{W}_{\!\!\bm{l}}$}

We now need to distinguish the true QCD expectation value
$\langle{\cal O}\rangle$ of any observable $\cal O$ from its
expectation value $\mbrak{\cal O}$ in
the theory with the modified quark determinant.
Only the latter can be estimated directly using
the representative ensembles of fields generated by the 
HMC algorithm, while the first is obtained through
\begin{equation}
  \langle{\cal O}\rangle=\frac{\mbrak{{\cal O}\,W_l}}{\mbrak{W_l}}.
  \label{rew}
\end{equation}
Evidently, for the reweighting (\ref{rew})
to work out in practice,
the statistical fluctuations of $W_l$ 
must be fairly small.

Whether this condition can be met on large lattices is
unclear since
\begin{equation}
   W_l=\rme^{-X_l},
   \qquad
   X_l=\int_0^{\mu^2}\rmd s_1\ldots\rmd s_l\, 
   \Tr\left\{(D^\dagger D+s_1+\ldots+s_l)^{-l}\right\},
\end{equation}
is the exponential of an extensive quantity $X_l$. 
In particular, using the moment-cumulant transformation one
can show that
\begin{equation}
   \frac{\mbrak{W_l^2}}{\mbrak{W_l}^2}=
   \nbrak{W_l}\nbrak{W_l{}^{-1}}=
   \exp\left\{\sum_{n=1}^{\infty}\frac{2}{(2n)!}
   \cbrak{X_l^{2n}}\right\},
   \label{cum_exp}
\end{equation}
where $\cbrak{X_l^{2n}}$ denotes the connected part of $\nbrak{X_l^{2n}}$. 
The fluctuations of the reweighting factor thus grow exponentially with 
the lattice volume $V$ 
and may therefore rapidly become too large 
when $V$ is increased. 
However, 
as we shall see in the following sections,
there are important mechanisms that suppress 
the fluctuations to the extent that 
determinant reweighting becomes a viable method in a useful range of 
parameters.

\section{Suppression of the high modes}

Determinant reweighting is intended for use at quark masses
close to or below the range of stability of the HMC algorithm.
The values of the renormalized
light-quark mass where reweighting will be 
applied are therefore expected to be smaller than $20$ MeV or so%
\kern1pt\footnote{
The renormalized masses are $\mr=\ZA m/\ZP$ and
$\mur=\mu/\ZP$, where $\ZP$ denotes the renormalization constant 
of the isovector pseudo-scalar density. 
The median $\nub$ of the 
distribution of the $n$'th eigenvalue $\nu_n$ of $(D^{\dagger}D)^{1/2}$
is similarly renormalized through $\nubr=\nub/\ZP$.
Values of these quantities quoted in physical units refer to the 
$\msbar$ renormalization scheme at $2$ GeV.}. 
At these quark masses, the low end of the spectrum of 
$(D^{\dagger}D)^{1/2}$ starts at about $\ZA m$
and has an approximately constant density from there up to
eigenvalues $\nubr$ of at least 100 MeV. 
In the following,
the associated eigenmodes of $D^{\dagger}D$
will be referred to as the ``low modes'' of the Dirac operator
and all other eigenmodes as the ``high modes''.
As explained in the next
section, the mass parameter $\mu$ will, in the cases of interest,
be less than, say, $2\kern1pt\ZA m$ and therefore always
well below the high eigenvalues of the Dirac operator.

The reweighting factors $W_l$ have been chosen so that 
the eigenvalues $\nu^2$ of $D^{\dagger}D$ larger
than $\mu^2$ make a monotonically decreasing contribution of 
order $\mu^{2l}/\nu^{2l}$ to $X_l$ (see Table~1).
Power counting then shows that the expectation values 
$\cbrak{X_l^{2n}}$ are ultraviolet convergent except for
$\cbrak{X_1^{2}}$ which diverges logarithmically.
The high modes of the Dirac operator thus 
contribute a term proportional to
$\mu^{4l}V$ to the fluctuations of $W_l$, with a proportionality
constant that diverges at most logarithmically in the continuum 
limit.

On current lattices the product $\mur^4V$
is usually much smaller than $1$. For lattices of
size $2L^4$ with $L\leq 4$ fm, for example, and if 
$\mur\leq20$ MeV,
one obtains $\mur^4V\leq0.054$. The high-mode
contribution to the statistical fluctuations of the 
reweighting factors $W_l$ thus tends to be strongly
suppressed, particularly so in the case of $W_2$, where
there is a second suppression factor proportional to 
$\mur^4$.

\section{Fluctuations of the low eigenvalues}

It is still not excluded, however, that 
the reweighting factors $W_l$ receive wildly fluctuating
contributions from the low modes of the Dirac operator.
There is no obvious suppression mechanism in this case,
and since the number of low modes grows proportionally to $V$, 
it seems likely that determinant reweighting will, in practice,
be limited to small lattices.

\begin{figure}
\centering
\epsfig{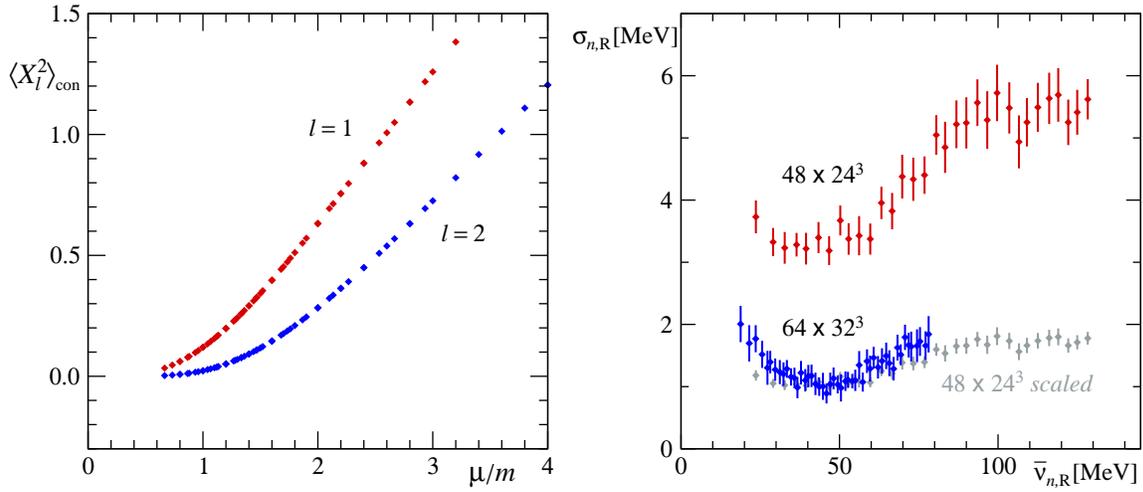}
\caption{
Values of $\cbrak{X_l^2}$ computed in random matrix theory
at $m=5,\ldots,30$ MeV
and $\mu=20,\ldots,44$ MeV, assuming 
$\Sigma=(250\;\MeV)^3$ and 
$V=(4.5\;\fm)^4$ (plot on the left). The plot on the right
shows the widths of the distributions
of the first $32$ ($48$) eigenvalues of $(D^{\dagger}D)^{1/2}$ on 
a lattice of size $48\times24^3$ ($64\times32^3$).
In both cases 
the lattice spacing and the renormalized 
sea-quark mass are approximately equal to 
$0.08$ fm and $25$ MeV respectively. The grey points labelled
``$48\times24^3$ scaled'' are
the $48\times24^3$ data scaled
by the ratio $(24/32)^4$ of the lattice volumes.
}
\label{fig1}
\end{figure}

In order to get some insight into the problem,
we worked out the leading term $\cbrak{X_l^2}$ of the 
cumulant expansion (\ref{cum_exp}) in the standard
two-flavour chiral random matrix theory \cite{ChRMT}.
In this theory, $\cbrak{X_l^2}$ can be expressed
through the spectral density of the Dirac operator and the 
spectral $2$-point correlation function,
both of which are known analytically \cite{ChRMT,DamgaardNishigaki}.
An integration over the spectral parameters is then still required, 
but since the integrands are non-singular, it is straightforward
to evaluate the integrals numerically.

In random matrix theory, 
$\cbrak{X_l^2}$ is a well-defined function of the
dimensionless combinations
$m\Sigma V$ and $\mu\Sigma V$, where $\Sigma$ denotes the 
quark condensate in the chiral limit.
To a very good approximation, we however found that
$\cbrak{X_l^2}$ only depends on the ratio of these parameters
(see Figure~1). Random matrix theory thus suggests that 
the contribution of the low modes 
to the fluctuations of the reweighting factors \emph{does not
change significantly with the volume $V$}. Moreover, as can be
seen from Figure~1, the contribution is actually quite small up
to values of $\mu/m$ equal to $2$ or so (note that $\ZA=1$ in 
random matrix theory and $\mu/m$ therefore corresponds to
$\mur/\mr$ in lattice QCD).

The outcome of our calculations in random matrix theory
can be explained by 
noting that random matrices have a fairly rigid spectrum,
i.e.~with high probability, the low eigenvalues are practically
unchanged from one random matrix to another.
The rigidity of the spectrum may well be related to
the fact that the Vandermonde determinant,
which appears in the joint distribution of the 
eigenvalues, gives rise to a repulsive force between
neighbouring eigenvalues. In any case, since the 
fluctuations of the low eigenvalues are of order $(\Sigma V)^{-1}$ 
and since there are $\rmO(V)$ eigenvalues, 
their contribution to the fluctuations of the reweighting factor $W_l$
will obviously 
remain bounded at large $V$.

In lattice QCD with Wilson quarks, chiral symmetry is not exactly
preserved and it is therefore not guaranteed that 
the eigenvalues of the lattice Dirac operator behave in 
the same way. Numerical studies of the 
O($a$)-improved two-flavour theory however suggest that
the widths of the eigenvalue distributions scale roughly 
like $1/V$, as in random matrix theory, except for the
widths measured close to the threshold of the spectrum
(see Figure~1)\kern1pt%
\footnote{The representative ensembles of gauge-field configurations
used for our numerical studies were generated  by 
the authors of ref.~\cite{CERN_TOVII} and were made available
to us through the 
\href{https://twiki.cern.ch/twiki/bin/view/CLS/WebHome}{CLS} community 
effort.}.
In the Wilson theory,
the contribution of the low modes to the fluctuations of the
reweighting factor is therefore slowly increasing  
with the lattice size, an effect that
is likely to become smaller as one moves closer to the continuum limit.

\section{Computation of the reweighting factors}

An exact calculation of the reweighting factors $W_l$ is normally
not possible and actually not required. 
Stochastic estimators can be used instead, or perhaps some combination
of a stochastic estimator and a (not necessarily exact) 
projector to the few lowest modes of the Dirac operator.
Here we only consider the most obvious choice,
where a set $\eta_k(x)$, 
$k=1,\ldots,\Neta$, of pseudo-fermion fields with action 
\begin{equation}
  S_{\eta}=\sum_{k=1}^{\Neta}\left(\eta_k,\eta_k\right)
\end{equation}
is added to the theory and the reweighting factor $W_l$ is replaced by
\begin{equation}
  W_{l,\Neta}=\frac{1}{\Neta}\sum_{k=1}^{\Neta}
  \exp\left\{\left(\eta_k,\left[1-w_l(D^{\dagger}D)^{-1}\right]\eta_k\right)
  \right\}.
  \label{srew}
\end{equation}
The simulation then proceeds as before and
the reweighting factor $W_{l,\Neta}$ is calculated according
to eq.~(\ref{srew}), using, for each gauge field, 
$\Neta$ randomly chosen pseudo-fermion fields.
This procedure is 
correct for any $\Neta\geq1$, but it pays to set $\Neta$  
to values significantly larger than $1$, because
the variance of $W_{l,\Neta}$ decreases
when $\Neta$ is increased (and eventually converges to the 
variance of $W_l$).

\begin{figure}
\centering
\epsfig{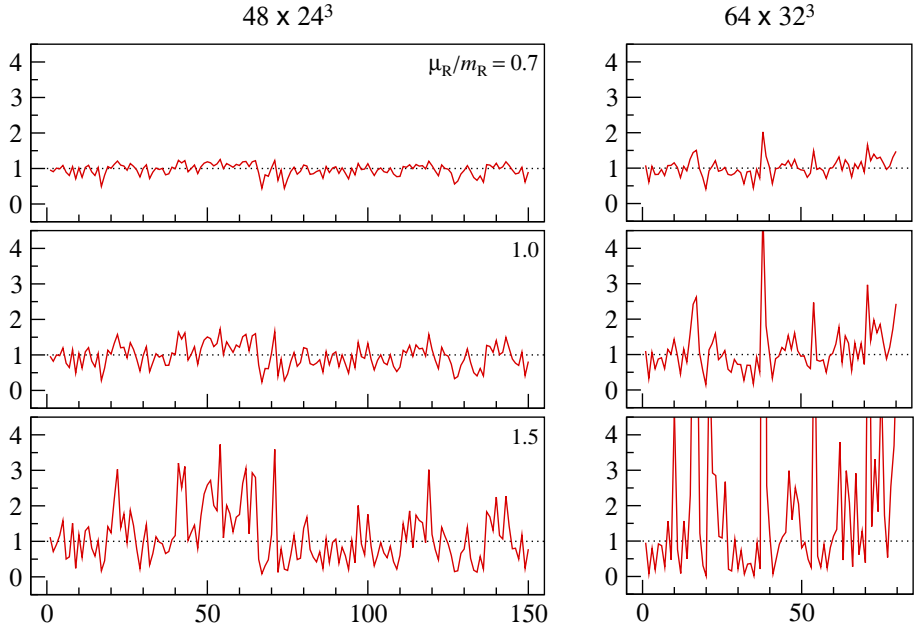}
\caption{
Reweighting factor $W_{2,24}$ normalized by its median
at $\mr\simeq25$ MeV and $\mur/\mr=0.7,1.0,1.5$ (from
top to bottom),
calculated for a set of
independent gauge-field configurations on a $48\times24^3$ (plots on the
left) and a $64\times32^3$ lattice (plots on the right)
with spacing $a\simeq0.08$ fm.
}
\label{fig2}
\end{figure}

For illustration the Monte Carlo time series of $W_{2,24}$ calculated
on the lattices previously considered are 
plotted in Figure~2. In all these cases, little would be gained by 
choosing more pseudo-fermion fields or by 
separating the lowest modes of the Dirac operator
(such a mode separation may, however, be required
at smaller quark masses).

Figure~2 also shows that the fluctuations of 
$W_{2,24}$ increase with the lattice size and that
they are quite sensitive to the value of $\mur/\mr$. 
In particular, by decreasing the latter, the fluctuations
are quickly reduced to acceptable levels on both lattices.
The fluctuations of $W_{1,24}$ at $\mur/\mr=0.7$ and $1.0$ are, 
incidentally, practically
the same as those of $W_{2,24}$ at $\mur/\mr=1.0$ and $1.5$, respectively. 
On the lattices considered
and before the performance of the HMC part of the algorithm is determined,
it is therefore not clear whether the first or the 
second factorisation of the quark determinant is preferable.

\section{Conclusions}

In this report we showed that determinant reweighting is 
likely to work out in lattice QCD if a factorisation 
of the quark determinant is chosen where the 
high-mode contribution
to the reweighting factor is sufficiently suppressed.
Somewhat surprisingly to us, the statistical 
fluctuations of the reweighting factor then do 
not grow rapidly with the volume of the lattice, 
a property that can be traced back to 
the rigidity of the spectrum of the low eigenvalues
of the Dirac operator.

We are now quite confident that simulations of the Wilson
theory at very small quark masses can be stabilised
in this way, but still need to prove this
by actually performing such simulations 
using the proposed determinant factorisations.

\end{document}